\documentclass[11pt]{article}
\usepackage{graphicx}

\begin{document}

\begin{center}

{\Large \bf The O(3,2) Symmetry derivable from the \\[2mm]
 Poincar\'e Sphere} \\
\vspace{3mm}
Y. S. Kim\\
Department of Physics, University of Maryland,\\
College Park, MD 20742, U.S.A. email: yskim@umd.edu

\end{center}

\vspace{3ex}

\begin{abstract}
Henri Poincar\'e formulated the mathematics of the Lorentz transformations,
known as the Poincar\'e group.  He also formulated the Poincar\'e sphere
for polarization optics.  It is noted that his sphere contains the symmetry
of the Lorentz group applicable to the momentum-energy four-vector of
a particle in the Lorentz-covariant world.  Since the particle mass is a
Lorentz-invariant quantity, the Lorentz group does not allow its variations.
However, the Poincar\'e sphere contains the symmetry corresponding to the
mass variation, leading to the $O(3,2)$ symmetry.  An illustrative
calculation is given.

\end{abstract}

\vspace{20mm}

\begin{flushleft}
Included in the Nova Editorial Book: Relativity, Gravitation, Cosmology:
Foundations, edited by Valeriy Dvoeglazov (2015).

\end{flushleft}

\newpage
\section{Introduction}\label{intro}
The Poincar\'e sphere is a mathematical instrument for studying polarization
of light waves~\cite{born80,bross98}.  This sphere contains the symmetry of
the Lorentz group~\cite{hkn97}.  In addition, the sphere allows
us to extend the $O(3,1)$ symmetry of the Lorentz group to the $O(3,2)$
symmetry of the de Sitter group~\cite{bk06jpa,kn13symm,bk13book}.

\par

In the Lorentz-covariant world, the energy and momentum are combined into a
four-vector, and the particle mass remains invariant under Lorentz
transformations.  Thus, it is not possible to change the particle mass in
the Lorentzian world.  However, in the de Sitter space of $O(3,2)$, there
are two energy variables allowing two mass variables $m_1$ and $m_2$, which
can be written as \cite{bk06jpa}
\begin{equation}\label{mass01}
m_1 = m \cos\chi, \quad\mbox{and}\quad   m_2 = m \sin\chi,
\end{equation}
respectively, with
\begin{equation}\label{mass02}
m^2 = m_{1}^2 + m_{2}^2 .
\end{equation}
For a given momentum whose magnitude is $p$, the energy variables are
\begin{equation}\label{mass03}
E_1 = \sqrt{ m^2\cos^{2}\chi + p^2}, \quad\mbox{and}\quad
E_2 = \sqrt{ m^2\sin^{2}\chi + p^2} .
\end{equation}

\par
While the Lorentz group is originally formulated in terms of the
four-by-four matrices applicable to one time and three space coordinates,
it is possible to use two-by-two matrices to perform the same Lorentz
transformation~\cite{hkn97,naimark54}.  In this representation,
the four-vector takes the form of a two-by-two Hermitian matrix with four
elements.  The determinant of this momentum-energy matrix is the $(mass)^2$
of this determinant.  Indeed, the Lorentz-transformation in this
representation consists of determinant-preserving transformations.

\par
When Einstein was formulating his special relativity, he did not consider
internal space-time structures or symmetries of the particles.   It was not
until 1939 when Wigner considered the space-time symmetries applicable to
the internal space-time symmetries.  For this purpose, Wigner in 1939
considered the subgroups of the Lorentz group whose transformations leave
the four-momentum of a given particle invariant~\cite{wig39,knp86}.
These subgroups are called Wigner's little groups, and they define the
internal space-time symmetries of the particles.

\par
For a massive particle, the internal space-time symmetry is like the
three-dimensional rotation group leading to the particle spin.  For a
massless particle, the little group has a cylindrical symmetry with one
rotational and one translational degrees of freedom, corresponding to
the helicity and gauge transformation respectively~\cite{kiwi90jmp}.
These Wigner's symmetry problems can also be framed into the two-by-two
formulation of the Lorentz group~\cite{bk13book}.

\par
It is known that the four Stokes parameters are needed for the complete
description of the Poincar\'e sphere~\cite{hkn97,bk06jpa}.  These parameters
can also be placed into a two-by-two matrix.  It is noted that phase shifts,
rotations, and amplitude change  lead to determinant-preserving
transformations, just like in the case of Lorentz transformations.

\par
However, the determinant of the Stokes parameters becomes smaller as
the two transverse components loses their coherence.   Since the determinant
of the two-by-two four-momentum matrix is the $(mass)^2$, this decoherence
could play as an analogy for variations in the mass.  The purpose of
this paper is precisely to study this decoherence mechanism in detail.

\par
Sec.~\ref{pew}, it is shown possible to study Wigner's little groups using
the two-by-two representation.  Wigner's little groups dictate the internal
space-time symmetries of elementary particles.  They are the subgroups of
the Lorentz group whose transformations leave the four-momentum of the
given particle invariant~\cite{wig39,knp86}.
\par
In Sec.~\ref{poincs}, we first note that the same two-by-two matrices are
applicable to two-component Jones vectors and Stokes parameters, which allow
us to construct the Poincar\'e spheres.   It is shown that the radius of the
sphere depends on the degree of coherence between two transverse electric
components.  The radius is maximum when the system is fully coherent and is
minimum when the system is totally incoherent.
\par
In Sec.~\ref{o32}, it is noted that the variation of the determinant of the
Stokes parameters can be formulated in terms of the symmetry of the $O(3,2)$
group~\cite{bk06jpa}.  This allows us to study the extra-Lorentz symmetry which allows
variations of the particle mass.  We study in detail the mass variation while
the momentum is kept constant.  The energy takes different values when
the mass changes.

\section{Poincar\'e Group, Einstein, and Wigner}\label{pew}
The Lorentz group starts with a group of four-by-four matrices
performing Lorentz transformations on the Minkowskian vector
space of $(t, z, x, y),$ leaving the quantity
\begin{equation}\label{4vec02}
t^2 - z^2 - x^2 - y^2
\end{equation}
invariant.  It is possible to perform this transformation using two-by-two
representations~\cite{hkn97,knp86,naimark54}.  This mathematical aspect is
known as the $SL(2,c)$ as the universal covering group for the Lorentz group.

\par
In this two-by-two representation, we write the four-vector as a matrix
\begin{equation}
X = \pmatrix{t + z  &  x - iy \cr x + iy & t - z} .
\end{equation}
Then its determinant is precisely the quantity given in Eq.(\ref{4vec02}).
Thus the Lorentz transformation on this matrix is a determinant-preserving
transformation.  Let us consider the transformation matrix as
\begin{equation}\label{g22}
 G = \pmatrix{\alpha & \beta \cr \gamma & \delta}, \quad\mbox{and}\quad
  G^{\dagger} = \pmatrix{\alpha^* & \gamma^* \cr \beta^* & \delta^*} ,
\end{equation}
with
\begin{equation}
    \det{(G)} = 1.
\end{equation}
This matrix has six independent parameters.   The group of these $G$
matrices is known to be locally isomorphic to the group of four-by-four
matrices performing Lorentz transformations on the four-vector
$(t, z, x, y)$~\cite{hkn97,knp86,naimark54}.  For each matrix of this
two-by-two transformation, there is a four-by-four matrix performing
the corresponding Lorentz transformation on the four-dimensional Minkowskian
vector.

\par

The matrix $G$ is not a unitary matrix, because its Hermitian conjugate is
not always its inverse.   The group can have a unitary subgroup called
$SU(2)$ performing rotations on electron spins.  This $G$-matrix formalism
explained in detail by Naimark in 1954~\cite{naimark54}.
We shall see first that this representation is convenient for studying
the internal space-time symmetries of particles.  We shall then note that
this two-by-two representation is the natural language for the Stokes
parameters in polarization optics.

\par
With this point in mind, we can now consider the transformation
\begin{equation}\label{naim}
X' = G X G^{\dagger} .
\end{equation}
Since $G$ is not a unitary matrix, it is not a unitary transformation.
For this transformation, we have to deal with four complex numbers.  However,
for all practical purposes, we may work with two Hermitian matrices
\begin{equation}\label{herm11}
Z(\delta) = \pmatrix{e^{-i\phi/2} & 0 \cr 0 & e^{i\phi/2}},
\quad\mbox{and}\quad
R(\phi) = \pmatrix{\cos(\theta/2)  & -\sin(\theta/2) \cr
     \sin(\theta/2) & \cos(\theta/2)} ,
\end{equation}
plus one symmetric matrix
\begin{equation}\label{symm11}
B(\mu) = \pmatrix{e^{\mu/2} & 0 \cr 0 & e^{-\mu/2}} .
\end{equation}
The two Hermitian matrices in Eq.(\ref{herm11}) lead to rotations around
the $z$ and $y$ axes respectively.  The symmetric matrix of Eq.(\ref{symm11})
performs Lorentz boosts along the $z$ direction.
Repeated applications of these three matrices will lead to the most general
form of the $G$ matrix of Eq.(\ref{g22}) with six independent parameters.

\par

It was Einstein who defined the energy-momentum four vector, and showed
that it also has the same Lorentz-transformation law as the space-time
four-vector.  We write the energy-momentum four-vector as
\begin{equation}\label{momen11}
P = \pmatrix{E + p_z & p_x - ip_y \cr p_x + ip_y & E - p_z} ,
\end{equation}
with
\begin{equation}
\det{(P)} = E^2 - p_x^2 - p_y^2 - p_z^2,
\end{equation}
which means
\begin{equation}\label{mass07}
\det{{p}} = m^2,
\end{equation}
where $m$ is the particle mass.
\par
Now Einstein's transformation law can be written as
 \begin{equation}
 P' = G M G^{\dagger} ,
 \end{equation}
or explicitly
\begin{equation}\label{lt03}
\pmatrix{E' + p_z' & p_x' - ip_y' \cr p'_x + ip'_y & E' - p'_z}
 = \pmatrix{\alpha & \beta \cr \gamma & \delta}
  \pmatrix{E + p_z & p_x - ip_y \cr p_x + ip_y & E - p_z}
  \pmatrix{\alpha^* & \gamma^* \cr \beta^* & \delta^*} .
\end{equation}

\par
Later in 1939~\cite{wig39}, Wigner was interested in constructing subgroups
of the Lorentz group whose transformations leave a given four-momentum
invariant, and called these subsets ``little groups.'' Thus, Wigner's
little group consists of two-by-two matrices satisfying
\begin{equation}\label{wigcon}
P = W P W^{\dagger} .
\end{equation}
This two-by-two $W$ matrix is not an identity matrix, but tells about
internal space-time symmetry of the particle with a given energy-momentum
four-vector.  This aspect was not known when Einstein formulated his special
relativity in 1905.

\par

If its determinant is a positive number, the $P$ matrix can be
brought to the form
\begin{equation}\label{massive}
         P = \pmatrix{1 & 0 \cr 0 & 1},
\end{equation}
corresponding to a massive particle at rest.

\par
If the determinant if zero, we may write $P$ as
\begin{equation}\label{massless}
         P = \pmatrix{1 & 0 \cr 0 & 0} ,
\end{equation}
corresponding to a massless particle moving along the $z$ direction.
\par
For all three of the above cases, the rotation matrix $Z(\phi)$ of
Eq.(\ref{herm11}) will satisfy the Wigner condition of Eq.(\ref{wigcon}).
This matrix corresponds to rotations around the $z$ axis.

\begin{table}
\caption{Wigner's Little Groups.  The little groups are the subgroups of
the Lorentz group whose transformations leave the four-momentum of a given
particle invariant.  They thus define the internal space-time symmetries of
particles.  The four-momentum remains invariant under the rotation around it.
In addition, they remain invariant under the following transformations.  They
are different for massive and massless particles.}\label{tab11}
\begin{center}
\begin{tabular}{lcl}
\hline
\hline \\[0.5ex]
 Particle mass &  Four-momentum  &  Transform matrices \\[1.0ex]
\hline\\
massive  & $\pmatrix{1 & 0 \cr 0 & 1}$
&
$\pmatrix{\cos(\theta/2) & -\sin(\theta/2)\cr \sin(\theta/2) & \cos(\theta/2)}$
\\[4ex]
Massless  &
$\pmatrix{1 & 0 \cr 0 & 0}$
& $\pmatrix{1 & \gamma \cr 0 & 1}$
\\[4ex]
\hline
\hline\\[-0.8ex]
\end{tabular}
\end{center}
\end{table}

\par
For the massive particle with the four-momentum of Eq.(\ref{massive}),
the two-by-two rotation matrix $R(\theta)$ also leaves the $P$ matrix of
Eq.(\ref{massive}) invariant.  Together with the $Z(\phi)$ matrix, this
rotation matrix lead to the subgroup consisting of unitary subset of the
$G$ matrices.  The unitary subset of $G$ is $SU(2)$ corresponding to the
three-dimensional rotation group dictating the spin of the
particle~\cite{knp86}.
\par
For the massless case, the transformations with the triangular matrix of
the form
\begin{equation}
\pmatrix{1 & \gamma \cr 0 & 1}
\end{equation}
leaves the momentum matrix of Eq.(\ref{massless}) invariant.  The physics of
this matrix has a stormy history, and the variable $\gamma$ leads to gauge
transformation applicable to massless particles~\cite{kiwi90jmp,hks82}.
\par

Table~\ref{tab11} summarizes the transformation matrices for Wigner's
subgroups for massive and massless particles.  Of course, it is a
challenging problem to have one expression for both cases, and this
problem has been addressed in the literature~\cite{bk10jmo}.

\section{Geometry of the Poincar\'e Sphere}\label{poincs}
The geometry of the Poincar\'e sphere for polarization optics is determined
the four Stokes parameters.  In order to construct those parameters, we
have to start from the two-component Jones vector.
\par
In studying the polarized light propagating along the $z$ direction, the
traditional approach is to consider the $x$ and $y$ components of
the electric fields.  Their amplitude ratio and the phase difference
determine the state of polarization.  Thus, we can change the polarization
either by adjusting the amplitudes, by changing the relative phase,
or both.  For convenience, we call the optical device which changes
amplitudes an ``attenuator'' and the device which changes the relative
phase a ``phase shifter.''
\par
The traditional language for this two-component light is the Jones-matrix
formalism which is discussed in standard optics textbooks.
In this formalism, the above two components are combined into one column
matrix with the exponential form for the sinusoidal function
\begin{equation}\label{jvec11}
\pmatrix{\psi_1(z,t) \cr \psi_2(z,t)} =
\pmatrix{a \exp{\left\{i\left(kz - \omega t + \phi_{1}\right)\right\}}  \cr
b \exp{\left\{i\left(kz - \omega t + \phi_{2}\right)\right\}}} .
\end{equation}
This column matrix is called the Jones vector.  To this vector we can
apply the following two diagonal matrices.
\par
\begin{equation}\label{shif11}
Z(\phi) = \pmatrix{e^{-i\phi/2} & 0 \cr 0 & e^{i\phi/2}} ,  \qquad
B(\mu) = \pmatrix{e^{\mu/2} & 0 \cr 0 & e^{-\mu/2}} .
\end{equation}
which leads to a phase shift and a change in the amplitudes respectively.  The
polarization axis can rotate around the $z$ axis, and it can be carried out
by the rogation matrix
\begin{equation}\label{rot11}
R(\theta) = \pmatrix{\cos(\theta/2) & -\sin(\theta/2)
\cr \sin(\theta/2) & \cos(\theta/2)} .
\end{equation}
These two-by-two matrices perform transform clearly defined in optics, while
they play the same role in Lorentz transformations as noted in Sec.~\ref{pew}.
Their role in the two different fields of physics are tabulated in
Table~\ref{tab22}.

The physical instruments leading to these matrix operations are mentioned
in the literature~\cite{hkn97,bk13book}.  With these operations, we can
obtain the most general form given in Eq.(\ref{jvec11}) by applying the
matrix $B(\mu)$ of Eq.(\ref{shif11})
to
\begin{equation}\label{jvec33}
\pmatrix{\psi_1(z,t) \cr \psi_2(z,t)} =
\pmatrix{a \exp{\left\{i(kz - \omega t - \phi/2)\right\}}  \cr
a \exp{\left\{i(kz - \omega t + \phi/2)\right\}}} .
\end{equation}
Here both components have the same amplitude.
\par
However, the Jones vector alone cannot tell whether the two components are
coherent with each other.  In order to address this important degree of
freedom, we use the coherency matrix defined as~\cite{born80,bross98}
\begin{equation}\label{cocy11}
C = \pmatrix{S_{11} & S_{12} \cr S_{21} & S_{22}},
\end{equation}
with
\begin{equation}
<\psi_{i}^* \psi_{j}> = \frac{1}{T} \int_{0}^{T}\psi_{i}^* (t + \tau) \psi_{j}(t) dt,
\end{equation}
where $T$ is for a sufficiently long time interval, is much larger
than $\tau$.  Then, those four elements become~\cite{hkn97}
\begin{eqnarray}
&{}& S_{11} = <\psi_{1}^{*}\psi_{1}> =a^2  , \qquad
S_{12} = <\psi_{1}^{*}\psi_{2}> = a^2~e^{-(\sigma +i\phi)} , \nonumber \\[1ex]
&{}& S_{21} = <\psi_{2}^{*}\psi_{1}> = a^2~e^{-(\sigma -i\phi)} ,  \qquad
S_{22} = <\psi_{2}^{*}\psi_{2}>  = a^2 .
\end{eqnarray}
The diagonal elements are the absolute values of $\psi_1$ and $\psi_2$
respectively.  The off-diagonal elements could be smaller than the product
of $\psi_1$ and $\psi_2$, if the two transverse components  are not
completely coherent.  The $\sigma$ parameter specifies the degree of
coherence.
\par

\begin{table}[ht]
\caption{Polarization optics and special relativity sharing the same
mathematics.  Each matrix has its clear role in both optics and relativity.
The determinant of the Stokes or the four-momentum matrix remains invariant
under Lorentz transformations.  It is interesting to note that the
decoherency parameter (least fundamental) in optics corresponds to the mass
(most fundamental) in particle physics.}\label{tab22}
\vspace{2mm}
\begin{center}
\begin{tabular}{lcl}
\hline
\hline \\[0.5ex]
 Polarization Optics & Transformation Matrix  &  Particle Symmetry \\[1.0ex]
\hline \\
Phase shift $\phi$  &
$\pmatrix{e^{-i\phi/2} & 0\cr 0 & e^{i\phi/2}}$
&  Rotation around $z$
\\[4ex]
Rotation around $z$  &
$\pmatrix{\cos(\theta/2) & -\sin(\theta/2)\cr \sin(\theta/2) & \cos(\theta/2)}$
&  Rotation around  $y$
\\[4ex]
Squeeze along $x$ and $y$  &
$\pmatrix{e^{\mu/2} & 0\cr 0 & e^{-\mu/2}}$
&  Boost along $z$
\\[4ex]
$(a)^{4} \left(1 - e^{-\sigma}\right)$  & Determinant &  (mass)$^2$
\\[4ex]
\hline
\hline\\[-0.8ex]
\end{tabular}
\end{center}
\end{table}

If we start with the Jones vector of the form of Eq.(\ref{jvec11}), the
coherency matrix becomes
\begin{equation}\label{cocy22}
C = a^4\pmatrix{1 & e^{-(\sigma + i\phi)} \cr e^{-(\sigma - i\phi)} & 1} .
\end{equation}
This is a Hermitian matrix and can be diagonalized to
\begin{equation}\label{cocy55}
D = a^4\pmatrix{1 + e^{-\sigma}  & 0 \cr 0 &  1 - e^{-\sigma} } .
\end{equation}

\par
For the purpose of studying the Poincar\'e sphere, it is more convenient to
make the following linear combinations.
\begin{eqnarray}\label{stokes11}
&{}& S_{0} = \frac{S_{11} + S_{22}}{\sqrt{2}},  \qquad
    S_{3} = \frac{S_{11} - S_{22}}{\sqrt{2}},    \nonumber \\[2ex]
&{}& S_{1} = \frac{S_{12} + S_{21}}{\sqrt{2}}, \qquad
S_{2} = \frac{S_{12} - S_{21}}{\sqrt{2} i}.
\end{eqnarray}
These four parameters are called Stokes parameters, and they are like a
Minkowskian four-vector which are Lorentz-transformed by the four-by-four
matrices constructed from the two-by-two matrices applicable to the
coherency matrix~\cite{hkn97}.
\par

We now have the four-vector $\left(S_0, S_3, S_1, S_2\right)$, and the
sphere defined in the three-dimensional space of
$\left(S_0, S_3, S_1, S_2\right)$ is called the Poincar\'e sphere.
If we start from the Jones vector of Eq.(\ref{jvec33}) with the same
amplitude for both components, $S_{11} = S_{22}$, and thus $S_{3} = 0.$
The Poincar\'e sphere becomes a two-dimensional circle.   The radius of
this circle is
\begin{equation}
 R = \sqrt{S_{1}^2 + S_{2}^2}.
\end{equation}
This radius takes its maximum value $S_{0}$ when the system is completely
coherent with $\sigma = 0$, and it vanishes when the system is totally
incoherent with $\sigma = \infty.$  Thus, $R$ can be written as
\begin{equation}
 R = S_{0} e^{-\sigma} .
\end{equation}
\par
Let us go back to the four-momentum matrix of Eq.(\ref{momen11}).  Its
determinant is $m^{2}$ and remains invariant under Lorentz transformations
defined by the Hermitian matrices of Eq.(\ref{herm11}) and the symmetric
matrix of Eq.(\ref{symm11}).  Likewise, the determinant of the coherency
matrix of Eq.\ref{cocy22} should also remain invariant.
The determinant in this case is
\begin{equation}
 S_0^2 - R^2 = a^{4} \left(1 - e^{-2\sigma}\right) .
\end{equation}
However, this quantity depends on the $\sigma$ variable which measures
decoherency of the two transverse components.  This aspects is illustrated
in Table~\ref{tab22}.
\par
While the decoherency parameter is not fundamental and is influenced by
environment, it plays the same mathematical role as in the particle mass
which remains as the most fundamental quantity since Isaac Newton,
and even after Einstein.

\section{O(3,2) symmetry}\label{o32}
The group $O(3,2)$ is the Lorentz group applicable to a five-dimensional
space applicable to three space dimensions and two time dimensions.
Likewise, there are two energy variables, which lead to a five-component
vector
\begin{equation}\label{desi03}
\left(E_1, E_2, p\right) = \left(E_{1},  E_{2}, p_z, p_x, p_y \right) .
\end{equation}
In order to study this group, we have to use five-by-five matrices, but
we are interested in its subgroups. First of all, there is a
three-dimensional Euclidean space consisting of $p_z, p_x,$ and $p_y$,
to which the $O(3)$ rotation group is applicable, as in the case of the
$O(3,1)$ Lorentz group.

\par
\begin{figure}
\centerline{\includegraphics[scale=0.7]{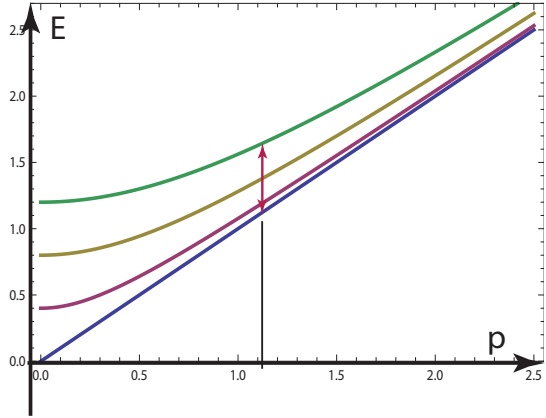}}
\caption{Energy-momentum hyperbolas for different values of the mass.  The
Lorentz group does not allow us to jump from one hyperbola to another, but
it is possible within the framework of the $O(3,2)$ de Sitter symmetry.
This figure illustrate the transition while the magnitude of the momentum
is kept constant.}\label{mvari}
\end{figure}

If the momentum is in the z direction, this five-vector becomes
\begin{equation}\label{desi05}
\left(E_1, E_2, p\right) = \left(E_{1},  E_{2}, p, 0, 0 \right).
\end{equation}
As for these two energy variables, they take the form
\begin{equation}\label{desi07}
E_{1} = \sqrt{p^2 + m^2 \cos^2\chi}, \quad\mbox{and}\quad
E_{2} = \sqrt{p^2 + m^2 \sin^2\chi} ,
\end{equation}
as given in Eq.(\ref{mass03}), and they maintain
\begin{equation}\label{desi09}
 E_{1}^2 + E_{2}^2  = m^{2} + 2p^{2},
\end{equation}
which remains constant for a fixed value of $p^2$.  There is thus a
rotational symmetry in the two-dimensional space of $E_{1}$ and $E_{2}$.
In this section, we are interested in this symmetry for a fixed value
of the momentum as described in Fig.~\ref{mvari}.

\par
For the present purpose, the most important subgroups are two Lorentz
subgroups applicable to the Minkowskian spaces of
\begin{equation}
\left(E_1, p, 0, 0\right), \quad\mbox{and}\quad \left(E_2, p, 0, 0\right) .
\end{equation}
Then, in the two-by-two matrix representation, these four-momenta
take the form
\begin{equation}\label{fvec21}
\pmatrix{E_1 + p & 0 \cr 0 & E_1 - p}, \quad\mbox{and}\quad
\pmatrix{E_2 + p & 0 \cr 0 & E_2 - p},
\end{equation}
with their determinant are $m^2\cos^2\chi$ and $m^2\sin^2\chi$
respectively.
With this understanding, we can now concentrate only on the matrix with
$E_{2}$.  For $\chi = 0$, we are dealing with the massless particle, while
the particle mass takes its maximum value of $m$.

\par
Indeed, this matrix is in the form of the diagonal matrix for the coherency
matrix given in Eq.(\ref{cocy55}).  Thus, we can study the property of
the four-vector matrix of Eq.(\ref{fvec21}) in terms of the coherency
matrix, whose determinant depends on the decoherency parameter $\sigma$.
Let us now take the ratios of the two diagonal elements for these matrices
and write
\begin{equation}
\frac{1 - e^{-\sigma}}{1 + e^{-\sigma}} =
\frac{\sqrt{p^2 + m^2\sin^{2}\chi} - p}{\sqrt{p^2 + m^2\sin^{2}\chi} + p} ,
\end{equation}
which becomes
\begin{equation}\label{fvec25}
    \frac{\tanh(\sigma/2)}{1 - \tanh^{2}(\sigma/2)} =
\left(\frac{m}{2p}\right)^2 (\sin\chi)^2 .
\end{equation}
The right and left sides of this equation consist of the variable of
the coherency matrix and that of the four-momentum respectively.

\par

\par

If $\sigma = 0$, the optical system is completely coherent, and this leads
to the zero particle mass with $\chi = 0.$  The resulting two-by-two
matrices are proportional to the four-momentum matrix for a massless
particle given in Eq.(\ref{massless})

\par

The right side reaches its maximum value of $(m/2p)^{2}$ when
$\chi = 90^{o}$, while the left side monotonically increases as $\sigma$
becomes larger.   It becomes infinite as $\sigma$ becomes infinite.  For
the left side, this is possible only for vanishing values of momentum.
The resulting two-by-two matrices are proportional to the four-momentum
matrix for a massive particle at rest given in Eq.(\ref{massive}).

\par
The variable $\sigma$ has a concrete physical interpretation in
polarization optics.  The variable $\chi$ cannot be explained in the world
where the particle mass remains invariant under Lorentz transformations.
However, this variable has its place in the $O(3,2)$-symmetric world.

\section*{Concluding Remarks}
In this report, it was noted first that the group of Lorentz transformations
can be formulated in terms of two-by-two matrices.  In this formalism, the
momentum four-vector can be written in the form of a two-by-two matrix.  This
two-by-two formalism can also be used for transformations of the coherency
matrix in polarization optics.  Thus, the set of four Stokes parameters is
like a Minkowskian four-vector subject to Lorentz transformations.  The
geometry of the Poincar\'e sphere can be extended to accommodate these
transformations.
\par
The radius of the Poincar\'e sphere depends on the degree of coherence
between the two transverse components of electric fields of the optical
beam.  If the system is completely coherent, the Stokes matrix is like
that for the four-momentum of a massless particle. When the system is
completely incoherent, the matrix corresponds to that for a massive
particle at rest.  The variation of the decoherence parameter corresponds
to the variation of the mass.
\par
This mass variation is not possible in the $O(3,1)$ Lorentzian world, but
is possible if the world is extended to that of the $O(3,2)$ de Sitter
symmetry.  In this paper, a concrete calculation is presented for the mass
variation with a fixed momentum in the de Sitter space.

\section*{Acknowledgments}
This report is in part based on a paper presented at the International
Conference ``Spins and Photonic Beams at Interface,'' honoring Academician
F. I. Fedorov held in Minsk, Belarus (2011).  I am grateful to professor
Sergei Kilin for inviting me to this conference.  I would also like
to thank Sibel Ba\c{s}kal and Marilyn Noz for their many years of
cooperation with me on this subject.  Finally, I thank Valeriy Dvoeglazov
for inviting me to submit this paper and for pointing out an error in one
of the rotation angles.

\end{document}